\documentclass[11pt]{article}
\usepackage{amssymb,latexsym}

\makeatletter
\@addtoreset{figure}{section}
\def\thefigure{\thesection.\@arabic\c@figure}
\def\fps@figure{h, t}
\@addtoreset{table}{bsection}
\def\thetable{\thesection.\@arabic\c@table}
\def\fps@table{h, t}
\@addtoreset{equation}{section}

\makeatother
\begin{document}

\title{\vskip-.5in
Hamilton's principle for quasigeostrophic motion
\footnote{LANL Report LA-UR-97-2205}}
\author{Darryl D. Holm \thanks{%
Research supported by US DOE} \\
Theoretical Division and Center for Nonlinear Studies \\
Los Alamos National Laboratory \\
Los Alamos, NM 87545 USA \\
dholm@lanl.gov \and Vladimir Zeitlin \\
Laboratoire de Meteorologie Dynamique\\
BP 99 tour 15-25, 5 etage\\
Universite P. et M. Curie\\
4, pl. Jussieu\\
75252 Paris Cedex 05 France\\
zeitlin@lmd.ens.fr}
\date{PACS numbers: 47.10.+g, 47.20.Ky, 47.32. y}
\maketitle

\begin{abstract}
We show that the equation of quasigeostrophic (QG) potential vorticity
conservation in geophysical fluid dynamics follows from Hamilton's principle
for stationary variations of an action for geodesic motion in the $f$- plane 
case or its prolongation in the $\beta$ - plane case. This implies a new
momentum equation and an associated Kelvin circulation theorem for QG
motion. We treat the barotropic and two-layer baroclinic cases, as well as
the continuously stratified case.

\vspace{0.25in}
\footnoterule
\noindent
{\it To appear in Physics of Fluids}

\end{abstract}


\pagestyle{myheadings}
\markright{ {\it  D. D. Holm and V. Zeitlin
--- Hamilton's principle for QG}}


\section{Introduction}

The quasigeostrophic (QG) approximation is a basic tool for the analysis of
meso- and large-scale motion in geophysical and astrophysical fluid dynamics
\cite{Pedlosky87}. Physically, QG theory applies when the motion is nearly
in geostrophic balance, i.e., when pressure gradients nearly balance the
Coriolis force. Mathematically, in the simplest case of a barotropic fluid
in a domain ${\mathcal{D}}$ on the plane ${\mathbb R}^2$ with coordinates
$(x_1,x_2)$, QG dynamics in the
$\beta $-plane approximation is expressed by the following evolution
equation for the stream-function $\psi $ of the geostrophic fluid velocity $%
\mathbf{u}=\hat{\mbox{\boldmath{$z$}}}\times {\mbox{\boldmath{$\nabla$}}}%
\psi $
\begin{equation}
{\frac{\partial ({\Delta }\psi -{\mathcal{F}}\psi )}{\partial t}}+[\psi ,{%
\Delta }\psi ]
+\beta {\frac{\partial \psi}{\partial x_1}} =0,  \label{qgb}
\end{equation}
where $\partial /\partial t$ is the partial time derivative, ${\Delta }$ is
the planar Laplacian, ${\mathcal{F}}$ denotes rotational Froude number, $%
[a,b]\equiv {\frac{\partial (a,b)}{\partial (x_1,x_2)}}$ is the Jacobi
bracket (Jacobian) for functions $a$ and $b$ on ${\mathbb R}^2$ and $\beta $
is the gradient of the Coriolis parameter, $f$, taken as $f=f_0+\beta x_2$ in
the $\beta$-plane approximation, with constants $\beta $ and $f_0$. Neglecting
$\beta $ gives the $f$-plane approximation. The QG equation (\ref{qgb}) may be
derived from the basic equations of rotating shallow water flow by a proper
nondimensionalization and subsequent asymptotic expansion with the Froude 
number given by the square of the ratio of the characteristic scale of the motion
to the  deformation radius (cf. Pedlosky \cite{Pedlosky87}
and, for recent discussions, Allen and Holm \cite{AH} and Stegner and
Zeitlin \cite{sz}). Alternatively, equation (\ref{qgb}) may be written in terms of
the potential vorticity as
\begin{equation}
{\frac{\partial q}{\partial t}}+\mathbf{u}\cdot {\mbox{\boldmath{$\nabla$}}}%
q=0,\qquad q\equiv {\Delta }\psi -{\mathcal{F}}\psi +f\,.  
\label{qgb'}
\end{equation}
In this form of QG dynamics, the basic property of potential vorticity
conservation on geostrophic fluid parcels becomes obvious.

Much is known about the mathematical structure of the QG equations
(\ref{qgb}) and (%
\ref{qgb'}). In particular, they are Hamiltonian with a Lie-Poisson bracket
given in Weinstein \cite{AW83}
\begin{equation}
\{F,G\} = -\int dx_1dx_2\ q \Big[{\frac{{\delta} F}{{\delta} \mu}}, {\frac{{%
\delta} G}{{\delta} \mu}}\Big]\,,  \label{qg-lpb}
\end{equation}
where $\mu\equiv q-f$. In terms of the variable $\mu$, the Hamiltonian for
QG is expressed as
\begin{eqnarray}
H &=& {\frac{1 }{2}}\int dx_1dx_2\ \Big (|{\mbox{\boldmath{$\nabla$}}}
\psi|^2+{\mathcal{F}}\psi^2 \Big ) \nonumber \\
&=& {\frac{1 }{2}}\int dx_1dx_2\ \mu\ ({\mathcal{F}}-{\Delta})^{-1} \mu + {%
\frac{1 }{2}}\sum_i\oint_{{\gamma}_i}\psi\mathbf{u}\cdot d\mathbf{x}\,,
\label{qg-ham}
\end{eqnarray}
where ${\gamma}_i$ is the $i$-th connected component of the boundary. In what 
follows we shall limit ourselves by the cases where ${\mathcal{D}}$ is either a 
torus (periodic boundary conditions) or the whole plane ${\mathbb R}^2$ with 
decaying boundary conditions and, thus, the 
boundary
terms may be ignored and $\delta H/\delta \mu = \psi$.
Consequently, the Lie-Poisson bracket (\ref{qg-lpb}) gives, after integration
by parts, the dynamical equation for $\mu$,
\begin{equation}
\frac{\partial \mu}{ \partial t} = \{ \mu ,H \}
= -\,[\psi, q] = -\, {\bf u} \cdot
{\mbox{\boldmath{$\nabla$}}} q\,,
\label{qgb2}
\end{equation}
in agreement with the QG potential vorticity equation (\ref{qgb'}).
Casimirs of the Lie-Poisson bracket (\ref{qg-lpb}) are given by
$C_{\Phi}=\int dx_1dx_2\Phi(q)$ for an arbitrary function $\Phi$ and satisfy
$\{C_{\Phi},H\}=0$ for all Hamiltonians
$H(\mu)$. Level surfaces of the Casimirs $C_{\Phi}$ define coadjoint orbits of
the group of symplectic diffeomorphisms of the domain of the flow \cite{AW83},
\cite{zp}.

The QG approximation may be generalized to the baroclinic case of several
superimposed fluid layers \cite{Pedlosky87}. For the simplest two-layer model, the
equations of motion describe the potential vorticity conservation in each layer
\begin{equation}
{\frac{\partial q^{(i)}}{\partial t}}+\mathbf{u}^{(i)}\cdot{%
\mbox{\boldmath{$\nabla$}}} q^{(i)}=0, \qquad \mathbf{u}^{(i)}=\hat{%
\mbox{\boldmath{$z$}}}\times {\mbox{\boldmath{$\nabla$}}} \psi^{(i)}, 
\qquad i=1,2\;,  \label{qg2}
\end{equation}
where potential vorticities are defined as
\begin{equation}
q^{(i)}=d^{(i)}\Big({\Delta} \psi^{(i)} + (-1)^i {\mathcal{F}}^{(i)}( \psi^{(1)} -
\psi^{(2)}) + f\Big)  
\qquad i=1,2\;,  \label{pv2}
\end{equation}
and $d^{(i)}$ and ${\mathcal{F}}^{(i)}$ are the nondimensional layer depths
and rotational Froude numbers, respectively. In the two-layer case, these
quantities are related by ${\mathcal{F}}^{(1)}d^{(1)}={\mathcal{F}}%
^{(2)}d^{(2)}$. The symplectic structure of the $N$-layer QG equations is
well known \cite{HMRW}. The Poisson bracket for multilayer QG is
also Lie-Poisson of the type (\ref{qg-lpb}). The Hamiltonian in this case is
again quadratic and equal to the sum of the kinetic energies of the layers
and the so-called available potential energy. In the case of two layers the
Hamiltonian is given by
\begin{equation}
H = {\frac{1 }{2}}\int dx_1dx_2 \bigg(\sum_{i=1,2} d^{(i)}\ |{%
\mbox{\boldmath{$\nabla$}}} \psi^{(i)}|^2 +{\frac{1 }{2}}({\mathcal{F}}%
^{(1)}d^{(1)} + {\mathcal{F}}^{(2)}d^{(2)})\ |\psi^{(1)} -
\psi^{(2)}|^2\bigg)\,.  \label{qg2-ham}
\end{equation}

There is also a QG approximation for a continuously stratified baroclinic fluid
\cite{Pedlosky87}. In terms of the (positive) ``static stability" parameter
${\mathcal{S}}(z)$, the QG equation for a continuously stratified fluid is
\begin{equation}
{\frac{\partial}{\partial t}}\Big(\Delta\psi
-{\frac{\partial}{\partial z}}
{\frac{1}{{\mathcal{S}}(z)}}{\frac{\partial\psi}{\partial z}}\Big)
+ \mathbf{u}\cdot{\mbox{\boldmath{$\nabla$}}}
\Big(\Delta\psi
-{\frac{\partial}{\partial z}}
{\frac{1}{{\mathcal{S}}(z)}}{\frac{\partial\psi}{\partial z}}+f\Big) = 0\, .  
\label{qg-vort-eqn0}
\end{equation}
Again this is potential vorticity conservation on geostrophic fluid parcels.
In fact, this is the standard equation for oceanic synoptic motions 
\cite{Pedlosky87}. In this equation,
$\mathbf{u}=\hat{\mbox{\boldmath{$z$}}}\times {\mbox{\boldmath{$\nabla$}}}
\psi$ is the geostrophic velocity for a stratified fluid and the stream
function $\psi(x_1, x_2, z)$ depends on both horizontal and vertical
coordinates in the domain ${\mathcal{D}}\times[z_0,z_1]$. The Hamiltonian
formulation of baroclinic continuously stratified QG dynamics
(\ref{qg-vort-eqn0}) with a Lie-Poisson bracket of the type (\ref{qg-lpb}) is
given in Holm
\cite{Holm85}. The Hamiltonian in the case of appropriate boundary conditions in
the vertical direction is \cite{Holm85}
\begin{equation} 
H = {\frac{1}{2}}\int_{\mathcal{D}} dx_1dx_2\int_{z_0}^{z_1} dz 
\bigg[|{\mbox{\boldmath{$\nabla$}}}\psi|^2 
+ {\frac{1}{{\mathcal{S}}(z)}}
\bigg({\frac{\partial\psi}{\partial z}}\bigg)^2\ \bigg]\,,
\label{QG-strat-erg}
\end{equation} 
where the term 
${\frac{1}{2{\mathcal{S}}(z)}}({\frac{\partial\psi}{\partial z}})^2$ 
is the density of available potential energy in stratified QG theory
\cite{Pedlosky87}.

In this paper, we provide new variational principles for QG dynamics in several
cases, and formulate the corresponding momentum equations and Kelvin circulation
theorems in these cases. These variational principles for QG dynamics are
potentially useful in observational data assimilation and studies of the transport
and mixing properties of QG flows. Our derivations are based on the previously
mentioned fundamental relation between the QG equations and the group of symplectic
diffeomorphisms (area-preserving mappings) of the domain of flow ${\mathcal{D}}$.
Related ideas have recently been discussed from the Euler-Poincar\'e  viewpoint
in Holm, Marsden and Ratiu \cite{HMR97}. Our approach is similar to that of
Virasoro \cite{viras} who suggested using the generating function for symplectic
diffeomorphisms as a generalized canonical coordinate with the subsequent
construction of a Lagrangian via the Legendre transform of the known Hamiltonian
expressed in terms of this variable. However, such a procedure can be practically
realized only in case where the incompressibility constraint may be explicitly
solved, i.e., for diffeomorphisms that are sufficiently close to identity and,
thus, describe only the initial stage of dynamical evolution. We avoid
solving the incompressibility constraint and, instead, use the parameters of the
group of symplectic diffeomorphisms of the domain of the flow as dynamical
variables.

A frequently used variational formulation of two-dimensional incompressible flows
is the Clebsch formulation; see, e.g., Seliger and Whitham \cite{SelWhi}, 
Holm and Kupershmidt \cite{HK83}, Marsden and Weinstein \cite{MW83}, Salmon
\cite{Salmon} and references therein. This formulation can indeed be extended
successfully to derive the quasigeostrophic equations. However, we decline using
the Clebsch representation here because it is not universal. That is, not every
coadjoint orbit of a given group can be represented by a given set of Clebsch
variables. For a discussion of this point see, e.g., Zeitlin \cite{VZ1},
\cite{VZ2}, \cite{z2}; a more detailed analysis of Clebsch representations for the
case of symplectic diffeomorphisms will be given elsewhere. 

\section{Preliminaries: Variational principles for Euler dynamics}

We begin with the case of $f$-plane dynamics, for which
equations (\ref{qgb'}) and (\ref{qg2}) for potential vorticity convection
may be interpreted as geodesic equations on the group of symplectic
diffeomorphisms in the same way as for the classical 2d Euler equations
discussed in Arnold \cite{Ar}. The difference is in the choice of metric which is
determined by the relation between the stream-function and the (potential)
vorticity \cite{zp}, \cite{msz}. 
Without giving rigorous differential-geometric and
functional-analytic meanings to this interpretation we shall simply discuss it
heuristically, by using the guideline that in formulating a geodesic
variational principle on a manifold it is convenient to deal with two sets of
variables: those belonging to the linear (co-)tangent space (the dual of the Lie
algebra of planar divergenceless vector fields in our case); and those
belonging to the configuration manifold itself (the Lie group of
area-preserving mappings in our case). In fluid dynamics, this means using both
Eulerian and Lagrangian coordinates. See, e.g., Holm, Marsden and Ratiu
\cite{HMR97} for a discussion of the equivalence of these two descriptions from
the viewpoint of variational principles defined on Lie groups and Lie algebras.

\paragraph{Lagrangian description.}
As shown in Arnold \cite{Ar}, the square of the geodesic distance on the group of
symplectic diffeomorphisms is given by the kinetic energy of the fluid, and
by explicitly introducing the incompressibility constraint with the help of
a Lagrange multiplier (the pressure) one obtains the action principle for 2d
incompressible fluid dynamics
\begin{equation}
\delta S = 0, \quad \hbox{with} \quad
S = \int dt\int dl_1dl_2 \ \left ({\frac{1 }{2}} |\dot{\mathbf{x}}|^2
+ p (J - 1) \right )\,.  \label{S2}
\end{equation}
In this expression, ${\mathbf x}=(x_1,x_2)$, the variables
$x_1(l_1,l_2;t),x_2(l_1,l_2;t)$ are considered as coordinates of fluid particles
marked by Lagrangian labels $l_1,l_2$, and $J$ is the Jacobian of the map
$(l_1,l_2)\rightarrow(x_1,x_2)$, with $J^i_A={%
\partial x^i}/{\partial l^A}, \;J=\mathrm{det}(J^i_A)$. Finally, 
overdot $\dot{(\ \ )}$ denotes time derivative following the fluid parcels,
i.e., at constant Lagrangian labels. The standard Lagrangian equations of
motion for an ideal fluid  of a constant density are (see, e.g., Lamb \cite{Lamb})
\begin{equation}
\ddot{\mathbf{x}}+ {\frac{\partial p }{\partial \mathbf{x}}} = 0,\quad
J = 1\,.  \label{L2}
\end{equation}
These equations are found from Hamilton's principle upon varying $S$ in
$\mathbf{x}$, imposing the constraint $J={%
\frac{\partial(x_1,x_2)}{\partial(l_1,l_2)}} =1$ (area preservation) on the
map $(l_1,l_2)\rightarrow (x_1,x_2)$ and using the known
properties of Jacobian determinants
\begin{equation}
{\frac{\partial (a,b)}{\partial (l_1,l_2)}} = {\frac{\partial (a,b)}{%
\partial (x_1,x_2)}} {\frac{\partial (x_1,x_2)}{\partial (l_1,l_2)}} \quad%
\hbox{and}\quad {\frac{\partial (a,x_2)}{\partial (x_1,x_2)}} = {\frac{%
\partial a}{\partial x_1}}.  \label{jac}
\end{equation}
The action $S$ in equation (\ref{S2}) is quadratic, since the pressure term
vanishes on the constraint manifold. This is appropriate for a geodesic
principle. (For a discussion of this point, cf. \cite {dnf}.) As is
well-known,
equations (\ref{L2}) transform to the standard Euler equations for the
velocity field $\mathbf{u} = \dot{%
\mathbf{x}}$ upon using the formula
\begin{equation}
\ddot{\mathbf{x}} = {\frac{d \mathbf{u} }{d t}} \equiv
\Big({\frac{\partial }{\partial t}} + (\mathbf{u}
\cdot {\mbox{\boldmath{$\nabla$}}})\Big) \mathbf{u}\,.  \label{u}
\end{equation}
Note that the Lagrangian conservation of vorticity (Kelvin circulation
theorem in 2d) follows from the general structure of the equations (\ref{L2}%
). Eliminating the pressure in (\ref{L2}) gives
\begin{equation}
{\frac{\partial\ddot{x}_2}{\partial x_1}} 
-{\frac{\partial\ddot{x}_1}{\partial x_2}} = 0 
\,,  \label{vort}
\end{equation}
which is equivalent, in view of (\ref{jac}), to $\dot{\omega} = 0$, where
$\omega \equiv  
{\frac{ \partial \dot{x}_2 }{\partial x_1}}
- {\frac{\partial \dot{x}_1 }{\partial x_2}} 
= \hat{\mbox{\boldmath{$z$}}}\cdot\mathrm{curl}\mathbf{u}$
is the vertical component of vorticity.

\paragraph{Eulerian description.}
An alternative way to treat the action principle (\ref{S2}) is to consider
it from the very beginning in the Eulerian $(x_1,x_2)$ plane, as
\begin{equation}
S=\int dt\ L,
\quad 
L = \int dx_1dx_2\ \Big[{\frac{1 }{2}} D|\mathbf{u}|^2 -p(D-1)\Big]\,,
\label{lag2-u}
\end{equation}
and to perform variations at fixed $x_1,x_2$ and $t$ of the {\it inverse}
maps $(x_1,x_2)\rightarrow(l_1,l_2)$, viewed as Eulerian fields 
$l^A(x_1,x_2,t)$, $A=1,2$, which determine $\mathbf{u}$ and $D$ by (summing on
repeated indices)
\begin{equation}
{\frac{\partial l^A}{\partial t}}=-u^iD^A_i,
\quad D^A_i={\frac{\partial l^A}{\partial x^i}} \quad D=%
\mathrm{det}(D^A_i)\,.  
\label{lag-def}
\end{equation}
Variations  with respect to $l^A$ and $p$ of any action whose Lagrangian is of the
form $L(\mathbf{u},D;p)$ then yield \cite{HMR87}, \cite{DH96},
\begin{eqnarray}
{\delta} S &=& \int dt \int dxdy\ \Bigg\{ D (D^{-1})^i_A {\delta} l^A \Bigg[ {%
\frac{d}{dt}} {\frac{1}{D}}{\frac{{\delta} L}{{\delta} u^i}} + {\frac{1}{D}}{%
\frac{{\delta} L}{{\delta} u^j}}u^j_{,i} - \Big({\frac{{\delta} L}{{\delta} D%
}}\Big)_{,i}\,\Bigg]  \nonumber \\
&&\hspace{1in} -\, {\delta} p(D-1)\Bigg\}\,,  
\label{hpg}
\end{eqnarray}
where
$d/dt=\partial/\partial t+\mathbf{u}\cdot{\mbox{\boldmath{$\nabla$}}}$
is the material derivative.
Stationarity of the action $S(\mathbf{u},D;p)$ under variations in $l^A$ also
results naturally in Kelvin's circulation theorem for any action of this form,
since
\cite{HMR87}, \cite{DH96}
\begin{equation}
{\frac{d}{dt}}\oint_{{\gamma}(t)}{\frac{1}{D}}{\frac{{\delta} L}{{\delta}
\mathbf{u}}}\cdot d\mathbf{x} = \oint_{{\gamma}(t)} \Big[ {\frac{d}{dt}} {%
\frac{1}{D}}{\frac{{\delta} L}{{\delta} \mathbf{u}}} + {\frac{1}{D}}{\frac{{%
\delta} L}{{\delta} u^j}}{\mbox{\boldmath{$\nabla$}}} u^j\Big]\cdot d\mathbf{%
x} = \oint_{{\gamma}(t)}{\mbox{\boldmath{$\nabla$}}}\Big({\frac{{\delta} L}{{%
\delta} D}}\Big)\cdot d\mathbf{x}=0\, ,  \label{kel}
\end{equation}
where ${\gamma}(t)$ is any closed loop moving with the fluid velocity
$\mathbf{u}$. 
For the action (\ref{lag2-u}) we have the variational formulas
\begin{equation}
{\frac{1}{D}}{\frac{{\delta} L}{{\delta} \mathbf{u}}} = \mathbf{u} \,,
\qquad
{\frac{{\delta} L}{{\delta} D}} = {\frac{1 }{2}}|\mathbf{u}|^2 - p\,.
\label{vds}
\end{equation}
Hence, from the vanishing of the coefficient of ${\delta} l^A$ in (\ref{hpg}%
) we find the Euler equations for the velocity field,
\begin{equation}
{\frac{\partial}{\partial t}}\mathbf{u} - \mathbf{u}\times\mathrm{curl}%
\mathbf{u} +{\mbox{\boldmath{$\nabla$}}}\Big(p+{\frac{1 }{2}}|\mathbf{u}|^2%
\Big)=0\, ,  \label{qg-mot-eqn}
\end{equation}
upon using the fundamental vector identity of fluid dynamics,
\begin{equation}
(\mathbf{b}\cdot{\mbox{\boldmath{$\nabla$}}})\mathbf{a} + a_j{%
\mbox{\boldmath{$\nabla$}}} b^j =(\hbox{curl}\, \mathbf{a})\times\mathbf{b}
+ {\mbox{\boldmath{$\nabla$}}}(\mathbf{a}\cdot\mathbf{b})\, ,  \label{fvid}
\end{equation}
with, in this case, $\mathbf{a}={\frac{1}{D}}{\frac{{\delta} L}{{\delta}
\mathbf{u}}}$ and $\mathbf{b}=\mathbf{u}$.

Of course, geometrical considerations are not essential in formulating this
variational principle, because $\mathbf{x}(\mathbf{l},t)$ represents
the fluid particle positions and equation (\ref{L2}) is just Newton's second law
with a constraint. In fact, this variational principle was first written by
Lagrange himself \cite {Lagrange}. However, geometrical considerations are
sometimes helpful in its interpretation. For example, 2d incompressible fluid
dynamics in the presence of the spatially dependent Coriolis force 
$(f_0+\beta x_2)\mathbf{u}\times\hat{\mbox{\boldmath{$z$}}}$ 
may be obtained from this variational principle by a ``prolongation"
\cite{HMR97}, \cite{viras}, \cite{HMR87} of the momentum density (which is equal 
to the velocity in the present context) in either
(\ref{S2}), or (\ref{lag2-u}), i.e., by a shift
\begin{equation}
{\frac{{\delta} L}{{\delta} \mathbf{u}}}
\rightarrow{\frac{{\delta} L}{{\delta} \mathbf{u}}} + D\mathbf{R},
\quad \hbox{or, equivalently in this case,}\quad
\mathbf{u}\rightarrow\mathbf{u} + \mathbf{R},
\end{equation}
where
\begin{equation} 
\mathbf{R}=
\hat{\mbox{\boldmath{$y$}}}(x_1+\beta x_1x_2), \ \mathrm{so}\ \hat{
\mbox{\boldmath{$z$}}}\cdot\mathrm{curl}\, \mathbf{R} =f(x_1,x_2)=f_0+\beta
x_2.  \label{f-beta}
\end{equation}
Geometrically, this shift corresponds to a central extension of the Lie
algebra of symplectic diffeomorphisms \cite{zp}. The prolonged action may be 
rendered quadratic by introducing an auxiliary gauge field using the Kaluza-Klein 
construction \cite{HMR97}.	

\section{Variational principles for QG dynamics}

As in the case of classical Euler dynamics described in the previous section,
we shall use the parameters of the symplectic diffeomorphisms of the flow domain
as  basic dynamical variables for QG. Thus, at this level of approximation we
introduce ``quasigeostrophic particles" which move at the geostrophic velocity 
$\mathbf{u}=\hat{\mbox{\boldmath{$z$}}}\times{\mbox{\boldmath{$\nabla$}}}\psi$ 
and follow the quasigeostrophic fluid trajectories,FIn pre
 which result from a 
certain filtering of the primitive equations. These QG fluid
trajectories will be described as functions of Lagrangian coordinates given
by $x_1(l_1,l_2;t),x_2 (l_1,l_2;t)$ in ${\mathcal{D}}$. Such
quasigeostrophic particles are commonly used in the context of another
geophysical fluid dynamics approximation, namely semigeostrophy \cite{Hoskins}.

\paragraph{Lagrangian representation.}

To apply the geometric approach of the previous section to QG dynamics, we
consider a generalized quadratic action on the $f$-plane of the form (cf.
equation (\ref{S2})
\begin{eqnarray}
S &=& \int dt\int dl_1dl_2 dl^{\prime}_1dl^{\prime}_2 \ \Big[{\frac{1 }{2}}
\dot{\mathbf{x}}(l_1,l_2)\cdot {\mathcal{G}}(\mathbf{x}(l_1,l_2),\mathbf{x}%
(l^{\prime}_1,l^{\prime}_2))\dot{\mathbf{x}}(l^{\prime}_1,l^{\prime}_2)
\nonumber \\
&&+ \int dt\int dl_1dl_2 \ p (J - 1)\Big],  \label{Sqg}
\end{eqnarray}
where ${\mathcal{G}}(\mathbf{x},\mathbf{x}^{\prime})$ is the kernel (Green's
function) of a bounded self-adjoint scalar integral operator. Varying this action
with respect to $\mathbf{x}$ at fixed $\mathbf{l}$ and $t$ gives the following
(Eulerian) equations of motion
\begin{equation}
{\frac{d}{dt}} \Big(\int dx^{\prime}_1dx^{\prime}_2 {\mathcal{G}}(\mathbf{x},%
\mathbf{x}^{\prime}) \dot{\mathbf{x}^{\prime}}\Big) - \Big(\int
dx^{\prime}_1dx^{\prime}_2 {\frac{\partial}{\partial\mathbf{x}}}
{\mathcal{G}}(\mathbf{x},\mathbf{x}^{\prime})
\dot{x^{\prime}_i}\Big) \dot{x_i} + {%
\frac{\partial p}{\partial\mathbf{x}}} = 0\,,  \label{Lqg}
\end{equation}
where 
the change of variables from $\mathbf{l}$ to $\mathbf{x}$ is understood.
As before in (\ref{vort}), by eliminating the pressure and using identities (\ref
{jac}) we obtain a conservation law on parcels, $\dot{q}_g=0$, where
\begin{equation}
q_g \equiv -\ {\frac{\partial}{\partial x_2}}\int dx^{\prime}_1dx^{\prime}_2 {%
\mathcal{G}}(\mathbf{x},\mathbf{x}^{\prime})\dot{x^{\prime}}_1 +\, {\frac{%
\partial}{\partial x_1}}\int dx^{\prime}_1dx^{\prime}_2 {\mathcal{G}}(\mathbf%
{x},\mathbf{x}^{\prime})\dot{x^{\prime}}_2\,,  \label{potvort}
\end{equation}
This expression recovers the potential vorticity for QG in the f-plane,
i.e., $q_f={\Delta}\psi-{\mathcal{F}}\psi$ in equation (\ref{qgb'}), when
${\mathcal{G}}$ is chosen to represent $1-{\mathcal{F}}{\Delta}^{-1}$ on the
domain ${\mathcal{D}}$ and $\dot{\mathbf{x}}=\hat{\mbox{\boldmath{$z$}}}%
\times{\mbox{\boldmath{$\nabla$}}}\psi$ is used. The passage to the
$\beta$-plane case may be accomplished by the prolongation (\ref{f-beta}).

\paragraph{Eulerian representation.}

Prolonginging 
${\delta L}/{\delta \mathbf{u}}
\rightarrow{\delta L}/{\delta \mathbf{u}} + D\mathbf{R}$,
in the action
(\ref{Sqg}) written in the Eulerian velocity representation with integral
operator ${\mathcal{G}}=(1-{\mathcal{F}}{\Delta}^{-1})$ yields
$S=\int dt\ L$, with Lagrangian
\begin{equation}
L = \int dx_1dx_2\ \Big[{\frac{1 }{2}} D\mathbf{u}\cdot(1-{\mathcal{F}%
}{\Delta}^{-1})\mathbf{u} + D\mathbf{u}\cdot\mathbf{R}(x_1,x_2) - p(D-1)\Big]%
\,,  \label{lag-u}
\end{equation}
and produces the following variations at fixed $x_1,x_2,$ and $t$,
\begin{eqnarray}
{\frac{1}{D}}{\frac{{\delta} L}{{\delta} \mathbf{u}}} 
&=& \mathbf{R} + \mathbf{u} - {\frac{\mathcal{F}}{2}}{\Delta}^{-1}\mathbf{u}
- {\frac{\mathcal{F}}{2D}}{\Delta}^{-1}(D\mathbf{u})\,, 
\nonumber \\ 
{\frac{{\delta} L}{{\delta} D}} &=& {\frac{1
}{2}}\mathbf{u}\cdot(1-{%
\mathcal{F}}{\Delta}^{-1})\mathbf{u} + \mathbf{u}\cdot\mathbf{R} - p\,,
\nonumber \\ 
{\frac{{\delta} L}{{\delta} p}} &=& -D+1\,.
\label{vds1}
\end{eqnarray}
Hence, from equation (\ref{hpg}) for action principles of this type and the
fundamental vector identity (\ref{fvid}), we find the Eulerian QG motion equation,
\begin{equation}
{\frac{\partial}{\partial t}}(1-{\mathcal{F}}{\Delta}^{-1})\mathbf{u} -
\mathbf{u}\times\mathrm{curl}\Big((1-{\mathcal{F}}{\Delta}^{-1})\mathbf{u} +
\mathbf{R}\Big) +{\mbox{\boldmath{$\nabla$}}}\Big(p+{\frac{1 }{2}}\mathbf{u}%
\cdot(1-{\mathcal{F}}{\Delta}^{-1})\mathbf{u}\Big)=0\, ,  \label{qg-mot-eqn1}
\end{equation}
where we have used the symmetry of the Laplacian and substituted the constraint
$D=1$, imposed by varying $p$. The curl of this equation yields
\begin{equation}
\frac{\partial \mu}{\partial t} + \mathbf{u}\cdot{%
\mbox{\boldmath{$\nabla$}}}(\mu +f) + (\mu +f){%
\mbox{\boldmath{$\nabla$}}}\cdot\mathbf{u} = 0\, ,  \label{qg-vort-eqn}
\end{equation}
where $\mu=\hat{\mbox{\boldmath{$z$}}}\cdot\hbox{curl}(1-{%
\mathcal{F}}{\Delta}^{-1})\mathbf{u}$. The constraint $D=1$ implies ${%
\mbox{\boldmath{$\nabla$}}}\cdot\mathbf{u}=0$ (from the kinematic relation $%
\partial D/\partial t+{\mbox{\boldmath{$\nabla$}}}\cdot D\mathbf{u}=0$) and
when $\mathbf{u}=\hat{\mbox{\boldmath{$z$}}}\times{\mbox{\boldmath{$\nabla$}}%
} \psi$ is substituted, the equation for $\mu={\Delta} \psi-{%
\mathcal{F}} \psi$ yields the QG potential vorticity equation (\ref{qgb'}). The
QG version of the Kelvin circulation theorem follows from formula (\ref{kel}) as
\begin{equation}
{\frac{d}{dt}}\oint_{{\gamma}(t)}\Big((1-{\mathcal{F}}{\Delta}^{-1})\mathbf{u%
} + \mathbf{R}\Big)\cdot d\mathbf{x} =0\, ,  \label{kel.cor}
\end{equation}
for any closed loop ${\gamma}(t)$ moving with the QG fluid velocity ${\bf u}$
under the Eulerian motion equation (\ref{qg-mot-eqn1}). The QG motion equation
(\ref{qg-mot-eqn1}) is derived in Euler-Poincar\'e form via the Kaluza-Klein
construction in \cite{HMR97}.

For a geodesic stationary principle, the Lagrangian and the Hamiltonian coincide
\cite{dnf}. In our case, Legendre transforming the Lagrangian (\ref{lag-u}) gives
the Hamiltonian
\begin{equation} 
H = \int dx_1dx_2\ \Big[{\frac{1 }{2}} D\mathbf{u}\cdot(1-{\mathcal{F}%
}{\Delta}^{-1})\mathbf{u} + p(D-1)\Big]%
\,.  \label{ham-u}
\end{equation} 
Thus, the Lagrangian and the Hamiltonian do coincide for QG, in the
f-plane theory and on the constraint manifold $D=1$. Taking $%
\mathbf{u}=\hat{\mbox{\boldmath{$z$}}}\times {\mbox{\boldmath{$\nabla$}}}%
\psi $ and using the divergence theorem yields the Hamiltonian for QG in terms
of the stream function,
\begin{equation} 
H\Big|_{D=1} = {\frac{1}{2}}\int_{\mathcal{D}} dx_1dx_2\ \Big[
|{\mbox{\boldmath{$\nabla$}}}\psi|^2
+{\mathcal{F}}\psi^2 \Big]
\,, 
\label{ham-u1}
\end{equation} 
thus reproducing the QG Hamiltonian (\ref{qg-ham}) for the Lie-Poisson bracket
formulation discussed in the introduction.

\paragraph{The multi-layer QG equations.}

A similar variational program may be carried out in the case of the
multi-layer QG
equations. In what follows we shall confine our attention to the two-layer
case, the further extension to $N$ layers being straightforward. In order to build
an action principle, we should take into account that, geometrically,
two-layer QG dynamics corresponds to geodesic flow on the direct product
of two groups of symplectic diffeomorphisms \cite{msz}, each group being
related to the motion in one layer. We, thus introduce two sets of
Lagrangian fluid trajectories $\mathbf{x}$ and $\mathbf{y}$, the pressures $%
p^{(x,y)}$ in each layer and the corresponding integral operators
\begin{eqnarray}
{\mathcal{G}}^{(xx)} = d^{(x)} (1-{\mathcal{F}}^{(x)}{\Delta}^{-1}), \quad {%
\mathcal{G}}^{(yy)} = d^{(y)} (1-{\mathcal{F}}^{(y)}{\Delta}^{-1}),
\nonumber \\
{\mathcal{G}}^{(xy)} = {\frac{1 }{2}}({\mathcal{F}}^{(x)}d^{(x)}+{\mathcal{F}%
}^{(y)}d^{(y)}) {\Delta}^{-1}.  
\label{G2}
\end{eqnarray}
Here the superscripts $x,y$ correspond to $1,2$ in (\ref{qg2}),(\ref{pv2}).
The two-layer analog of the action (\ref{Sqg}) then may be written as
\begin{eqnarray}
S = \int dt\int dl_1dl_2 dl^{\prime}_1dl^{\prime}_2 \Big[{\frac{1 }{2}} \dot{%
\mathbf{x}}(l_1,l_2)\cdot{\mathcal{G}}^{(xx)}(\mathbf{x}(l_1,l_2), \mathbf{x}%
(l^{\prime}_1,l^{\prime}_2))\dot{\mathbf{x}}(l^{\prime}_1,l^{\prime}_2) +
\nonumber \\
{\frac{1 }{2}}\dot{\mathbf{x}}(l_1,l_2)\cdot{\mathcal{G}}^{(xy)}(\mathbf{x}%
(l_1,l_2), \mathbf{y}(l^{\prime}_1,l^{\prime}_2))\dot{\mathbf{y}}%
(l^{\prime}_1,l^{\prime}_2) +  \nonumber \\
{\frac{1 }{2}}\dot{\mathbf{y}}(l_1,l_2)\cdot{\mathcal{G}}^{(yy)} (\mathbf{y}%
(l_1,l_2),\mathbf{y}(l^{\prime}_1,l^{\prime}_2))\dot{\mathbf{y}}%
(l^{\prime}_1,l^{\prime}_2) \Big]  \nonumber \\
+ \int dt\int dl_1dl_2 \Big( p^{(x)} (J^{(x)} - 1) + p^{(y)} (%
 J^{(y)} - 1)\Big).\quad  \label{Sqg-2}
\end{eqnarray}
By performing the variations and eliminating the pressure as before we find
potential vorticity conservation in each layer. For example, in the $x$
layer we have
\begin{equation}
{\frac{d }{dt}} q^{(x)} = 0,  \label{pvcons2}
\end{equation}
where (using $D^{(x)}=1$ and $D^{(y)}=1$)
\begin{equation}
q^{(x)}=-{\frac{\partial U_1 }{\partial x_2}} 
+{\frac{\partial U_2 }{\partial x_1 }},\;\mathbf{U} 
= \int dx^{\prime}_1dx^{\prime}_2
{\mathcal{G}}^{(xx)}(\mathbf{x},\mathbf{x}^{\prime})
\dot{\mathbf{x}^{\prime}_1}+ \int dy^{\prime}_1dy^{\prime}_2
{\mathcal{G}}^{(xy)}(\mathbf{x},\mathbf{y} ^{\prime})
\dot{\mathbf{y}^{\prime}_1}.  
\label{potvort2}
\end{equation}
A similar expression holds with obvious changes in the $y$ layer. It is easy
to check that both these expressions coincide with the definitions 
(\ref{pv2}). Thus, two-layer QG dynamics is obtained from Hamilton's
principle for the action (\ref{Sqg-2}). Again, one could start by transforming the
action (\ref{Sqg-2}) into the Eulerian velocity representation by using two
sets of Lagrangian labels and arrive at the same multi-layer QG equations.

\paragraph{Continous vertical stratification.}

Consider an action of the following Eulerian form in three dimensions
$(x_1,x_2,z)$ with vertical coordinate $z$ in the interval $[z_0,z_1]$,
\begin{equation}
S = \int dt\int_{\mathcal{D}} dx_1dx_2\int_{z_0}^{z_1} dz\ 
\Big[{\frac{1 }{2}} D\mathbf{u}\cdot(1-{\mathcal{L}}(z)
{\Delta}^{-1})\mathbf{u} + D\mathbf{u}\cdot\mathbf{R}(x_1,x_2) - p(D-1)\Big].
\label{lag-uz}
\end{equation}
Let the operator ${\mathcal{L}}(z)$ depend only on $z$, 
so ${\mathcal{L}}(z)$ commutes with the planar Laplacian, $\Delta$. Hamilton's
principle for this action yields a motion equation corresponding to
(\ref{qg-mot-eqn1})
\begin{eqnarray}
{\frac{\partial}{\partial t}}
(1-{\mathcal{L}}(z){\Delta}^{-1})\mathbf{u} -
\mathbf{u}\times\mathrm{curl}\Big((1-{\mathcal{L}}(z){\Delta}^{-1})\mathbf{u} 
+ \mathbf{R}\Big) 
\nonumber\\
+\ {\mbox{\boldmath{$\nabla$}}}\Big(p+{\frac{1 }{2}}\mathbf{u}
\cdot(1-{\mathcal{L}}(z){\Delta}^{-1})\mathbf{u}\Big)=0\, ,  
\label{qg-mot-eqn2}
\end{eqnarray}
with ${\mbox{\boldmath{$\nabla$}}}
=(\partial/\partial x_1,\partial/\partial x_2)$, as
before. Thus, by equation (\ref{kel}), the Kelvin theorem for stratified QG
dynamics has the same form as the single layer Kelvin theorem (\ref{kel.cor}) on
each horizontal surface parameterized by the level $z=const$. Namely, cf. equation 
(\ref{kel.cor}),
\begin{equation}
{\frac{d}{dt}}\oint_{{\gamma}(t)}\Big((1-{\mathcal{L}}(z){\Delta}^{-1})\mathbf{u%
} + \mathbf{R}\Big)\cdot d\mathbf{x} =0\, ,  \label{kel.cor-strat}
\end{equation}
for each closed loop ${\gamma}(t)$ moving with the horizontal QG fluid velocity
${\bf u}$, governed by the Eulerian motion equation (\ref{qg-mot-eqn2}). This 
corresponds to ``pseudo-potential vorticity" conservation.
The curl of the stratified Eulerian QG motion
equation (\ref{qg-mot-eqn2}) yields, cf. equation (\ref{qg-vort-eqn}),
\begin{equation}
{\frac{\partial}{\partial t}}\Big(\Delta\psi-{\mathcal{L}}(z)\psi\Big)
+ \mathbf{u}\cdot{\mbox{\boldmath{$\nabla$}}}
\Big(\Delta\psi-{\mathcal{L}}(z)\psi+f\Big) = 0\, ,  
\label{qg-vort-eqn1}
\end{equation}
where
$\mathbf{u}
=\hat{\mbox{\boldmath{$z$}}}\times {\mbox{\boldmath{$\nabla$}}}\psi(x_1,x_2,z)$
is consistent with $D=1$ and the operator ${\mathcal{L}}(z)$ is still to be
specified. Choosing 
\begin{equation} 
{\mathcal{L}}(z)={\frac{\partial}{\partial z}}
{\frac{1}{{\mathcal{S}}(z)}}{\frac{\partial}{\partial z}},
\label{L-op}
\end{equation} 
which contains the (positive) ``static stability" parameter ${\mathcal{S}}(z)$,
produces
\begin{equation} 
\bigg[{\frac{\partial}{\partial t}}
+{\frac{\partial\psi}{\partial x_1}}{\frac{\partial}{\partial x_2}}
-{\frac{\partial\psi}{\partial x_2}}{\frac{\partial}{\partial x_1}}\bigg]
\bigg[{\frac{\partial^2\psi}{\partial x_1^2}}
+{\frac{\partial^2\psi}{\partial x_2^2}}
+{\frac{\partial}{\partial z}}\bigg(
{\frac{1}{{\mathcal{S}}(z)}}{\frac{\partial\psi}{\partial z}}\bigg)+f\bigg]=0\,.
\label{StratQG-eq}
\end{equation} 
This is the standard equation for oceanic synoptic motions,
cf. Pedlosky \cite{Pedlosky87}, p.364. 

Legendre transforming the Lagrangian in (\ref{lag-uz}) yields the (constrained)
Hamiltonian,
\begin{eqnarray}
H &=& \int dx_1dx_2dz
\Big[\mathbf{u}\cdot{\frac{\delta L}{\delta \mathbf{u}}}
-{\frac{1}{2}} D\mathbf{u}\cdot(1-{\mathcal{L}}(z)
{\Delta}^{-1})\mathbf{u} 
- D\mathbf{u}\cdot\mathbf{R}
+\ p(D-1)\Big]
\nonumber\\
&=& \int dx_1dx_2dz 
\Big[{\frac{1}{2}} D\mathbf{u}\cdot\Big(1-{\mathcal{L}}(z)
{\Delta}^{-1}\Big)\mathbf{u} 
+\ p(D-1)\Big]
\label{SQG-erg}\\
&=& \int dx_1dx_2dz 
\Big[{\frac{1}{2}} D|{\mbox{\boldmath{$\nabla$}}}\psi|^2
- {\frac{1}{2}} D {\mbox{\boldmath{$\nabla$}}}\psi
\cdot {\mathcal{L}}(z)\Delta^{-1}{\mbox{\boldmath{$\nabla$}}}\psi
+\ p(D-1)\Big]\,.
\nonumber
\end{eqnarray}
When this Hamiltonian is evaluated on the constraint manifold $D=1$, it
becomes the conserved energy,
\begin{equation} 
E = {\frac{1}{2}}\int dx_1dx_2dz 
\Big[|{\mbox{\boldmath{$\nabla$}}}\psi|^2 
- \psi{\mathcal{L}}(z)\psi\Big]\,,
\end{equation} 
after integrating by parts in $(x_1,x_2)$ and discarding the horizontal boundary
term. For the choice of the operator ${\mathcal{L}}$ in (\ref{L-op}), we then find
\begin{eqnarray} 
E &=& {\frac{1}{2}}\int dx_1dx_2dz 
\bigg[|{\mbox{\boldmath{$\nabla$}}}\psi|^2 
+ {\frac{1}{{\mathcal{S}}(z)}}
\bigg({\frac{\partial\psi}{\partial z}}\bigg)^2\ \bigg]
\nonumber\\
&&-\ {\frac{1}{2}}\int dx_1dx_2 \bigg[
{\frac{\psi}{{\mathcal{S}}(z)}}
{\frac{\partial\psi}{\partial z}}\bigg]^{z_1}_{z_0}\,,
\end{eqnarray} 
which is the conserved energy for ideal synoptic scale motions and the
Hamiltonian for the Lie-Poisson formulation of baroclinic QG theory, cf. equation
(\ref{QG-strat-erg}) and Holm \cite{Holm85}. A choice of ${\mathcal{L}}(z)$
extending that in equation (\ref{L-op}) yields the modified QG theory of White
\cite{White}, namely
\begin{equation} 
{\mathcal{L}}(z)=\Big({\frac{\partial}{\partial z}}+B\Big)
{\frac{1}{{\mathcal{S}}(z)}}\Big({\frac{\partial}{\partial z}}-B\Big) 
- {\mathcal{F}},
\label{L-op-mqg}
\end{equation}
where the definitions of the symbols may be obtained in \cite{White}. 
For a discussion of the Lie-Poisson Hamiltonian formulation of White's modified
QG theory, see Holm \cite{Holm91}.

\section{Discussion}

There are several potential applications of the QG variational principles
presented here. For example, by studying the second variation of the
action and, hence, the separation of the geodesics (cf. \cite{dnf}) one may
attack the problem of Lagrangian stability of QG flows and study dispersion
of the quasigeostrophic particles which may be considered as a test of accuracy 
and predictability of QG flows. Also, in the context of numerical
simulations, one may try to find proper discretizations directly at the level
of the action, as proposed in Holm et al. \cite{HKL85}. (See Marsden and
Wendlandt \cite{MW97}, \cite{WM} for recent advances in this direction.)
Finally, as QG dynamics is frequently used in assimilating atmospheric and
oceanic data, the variational principles presented here could be
useful in this domain, e.g., if optimization procedures were applied directly to
the QG action, cf. Bennett \cite{Bennett92}. To facilitate their utility, we have
expressed these QG variational principles in both Eulerian and Lagrangian
coordinates.

\bigskip

\paragraph{Acknowledgements.}
We thank Tudor Ratiu for many valuable discussions. This research was
performed while the authors were visiting the Isaac Newton Institute for
Mathematical Sciences at Cambridge University. We gratefully acknowledge
financial support by the program ``Mathematics in Atmosphere and Ocean
Dynamics" as well as the stimulating atmosphere at the Institute during our
stay there. We are also grateful to the anonymous referees for several useful
remarks.

\end{document}